\documentclass[a4]{article}
\usepackage{graphicx}
\begin{document}
\begin{center}
    {\large\bf Conservation laws and laser cooling of atoms}\\
    \vskip4mm
    Giuseppe Giuliani\\
    \vskip2mm
    Dipartimento di Fisica, Universit\`a di Pavia\\
    Via Bassi 6, 27100 Pavia\\
    \vskip2mm
    giuseppe.giuliani@unipv.it

\end{center}
\noindent
{\bf Abstract}.
The straightforward application of energy and linear momentum conservation to the absorption/emission of photons by atoms allows to establish the essential features of laser cooling of two levels atoms  at low laser intensities. The lowest attainable average kinetic energy of the atoms  depends on the ratio $\Gamma/E_R$ between the natural linewidth and the recoil energy and tends to $E_R$  as $\Gamma/E_R$ tends to zero.  This treatment, like the quantum mechanical ones,  is valid for any value of the ratio $\Gamma/E_R$ and contains  the  semiclassical theory of laser cooling as the limiting case in which $E_R\ll \Gamma$.

\vskip5mm\par\noindent
pacs {37.10.De 03.30.+p}
\section{Introduction}
 Starting from the pioneering works of  mid Seventies of last century, laser cooling of atoms has become a vast research field with many applications in physics, chemistry and biology.
    \par
    When an atom absorbs a photon, its kinetic energy is changed according to the laws of energy and linear momentum conservation. In general, but not always (section \ref{linassappsec}), if an atom flying against a photon absorbs it,   the kinetic energy of the atom decreases. Instead, if a photon is chasing the atom, the variation of the atom's kinetic energy due to the absorption of the photon is always positive.
    \par
     The proposal of cooling atoms with laser beams has been put forward by H\"{a}nsch and Schawlow in 1975 \cite{hs}: the idea was that of illuminating  the atoms with six laser beams (two opposite beams for each spatial dimension) red detuned with respect to an absorption line.
  \par
  Apparently,  there are two limits to the cooling process: the Doppler limit and the recoil limit. The Doppler limit is due to the natural width $\Gamma$ (Full Width at Half Maximum) of the atomic transition used:  since the first order Doppler shift is $\pm \Delta E (v/c)$ (where $\Delta E$ is the transition energy between the two atomic levels), when  $\Delta E (v/c)\approx \Gamma$, the photon may be absorbed (with significant probability) not only by atoms flying against the photon but also by atoms flying in the opposite direction, thus limiting the cooling process.
  Instead, the recoil limit is due to the fact that when the kinetic energy of the absorbing atom is of the same order of magnitude of its variation due to the absorption or emission of a photon no further cooling seems to be possible.
 \par
The theoretical treatment of laser cooling is not a simple one.  In the semi--classical approach, the atom is considered as a localized two--levels quantum system  and the light field is treated classically \cite{stenholm}.
 According to this theory, the lowest attainable average kinetic energy $<E_K>$ of the atoms is given by $\Gamma/4$. This result can not be valid in all conditions since it implies that $<E_K>\rightarrow 0$ as $\Gamma\rightarrow 0$. This physically unsound result is due to the assumption that the photon momentum  is negligible with respect to the atomic one. This approximation implies that the   kinetic energy of the atom is much larger than the recoil energy $E_R$. As we shall see, the semiclassical theory is applicable only when $\Gamma\gg E_R$.
 \par
 A  quantum mechanical treatment of the motion of a two levels atom under laser light, applicable for any value of the ratio between  the natural linewidth and the recoil energy, has been developed, among others,  by Winelnd and Itano \cite{wineland} and by Castin, Wallis and Dalibard \cite{castin}.
 \par
 The discovery  by Lett et al. that temperatures well below the Doppler limit can be achieved \cite{lett} inspired a re--formulation of the theoretical description: it was found that the multilevel nature of the atoms and the spatial variation  of the light field polarization can be exploited  for attaining temperatures close to the recoil limit \cite{dalibard}.
 \par
Temperatures below the recoil limit can be achieved by sophisticated procedures like the Velocity Selective Coherent Population Trapping \cite{dark} and the Stimulated Raman Cooling \cite{raman}.  The book by  Metcalf and  van der Straten  may be taken as a guide for exploring all these issues and the vast available literature \cite{metstra}.
\par
Stimulated by  these  developments, the search for ever lower temperatures has attracted  the attention of experimental and theoretical physicists. This notwithstanding, the theoretical reconsideration of laser cooling of two levels atoms maintains its importance, since it sheds light on  essential features of the cooling process.
\par
This paper is based on the idea that the application of  conservation laws to the absorption/emission of photons by atoms should yield the essential feature of laser cooling of two levels atoms at low laser intensities. We shall see that this is indeed the case,
\par
The conceptual framework of the following treatment is simple and the mathematics needed is limited to some algebraic manipulations. Therefore, this paper might be of some interest for university and high school teaching. It might also be of some value for researchers: being based on conservation principles, it sets down limiting conditions that should be met also by more sophisticated (and complicated) treatments like the quantum mechanical ones.
\par
 As we shall see in due course, the present treatment begins with the use of relativistic dynamics; however, given the small atoms' velocities of interest in the final steps of the laser cooling process, the relativistic dynamics will be approximated for small velocities. Of course, it is possible to use directly the  Newtonian mechanics (this may be a  choice for high school teaching). However, it is worth stressing the conceptual differences between the two approaches. Within the relativistic approach, the rest energy $Mc^2$ appears naturally with its physical meaning and the transition energy $\Delta E$ between two atomic levels, being the difference of two rest energies, is a relativistic invariant; instead, in the Newtonian treatment, the physical meaning of $Mc^2$, which enters the expression of the recoil energy, remains obscure.
\section{Outline of the paper}
An outline of the main steps of the paper will help in understanding  how the  various calculations contribute to the overall description.
 We shall deal only with energies: the  transition energy $\Delta E_M$, the recoil energy $E_R=\Delta E_M^2/2Mc^2$, the natural linewidth $\Gamma$ and the energy $E_{ph}$ of the absorbed/emitted photon. The transition energy $\Delta E_M$ is the one corresponding to the most probable value given by the Lorentzian shape of the natural line.
\par
In section \ref{basic}, we shall assemble the basic formulas, giving them the  most suitable form for the laser cooling process. Two dimensionless parameters are introduced:
 \begin{equation}\label{BT}
    B_T=\frac{E_R}{\Delta E_M}=\frac{\Delta E_M}{2Mc^2}
 \end{equation}
 and
 \begin{equation}\label{BD}
    B_D=\frac{\Gamma}{\Delta E_M}
 \end{equation}
 Their ratio $B_D/B_T$ is equal to $\Gamma/E_R$.
  A  third dimensionless parameter will enter the description:  the parameter $B_1=v_1/c$, where  $v_1$ is the norm of the atom's velocity  vector $\vec v_1$ before the absorption/emission of a photon. The use of dimensionless parameters simplify the formulas and allows an easy comparison of the orders of magnitude; however, when needed, the basic physical quantities  will be re--established.
\par
In section \ref{naturalsec}, the implications of  a non zero linewidth are discussed. In particular, it is shown that when an atom absorbs a counter--propagating photon its transition energy is $\Delta E^A$, while when the atom absorbs a co--propagating photon of the same energy its transition energy is $\Delta E^P=\Delta E^A(1-B_1)/(1+B_1)$.
\par
In section \ref{linear}, the basic formulas are approximated in the limit of small atoms' velocities: in this limit, only the first order Doppler effect is taken into account and the relativistic dynamics can be replaced by the Newtonian one.
 In order to identify this limit without ambiguity, we must take into account  that the  formulas  contain terms of the first or higher order in the $B$'s ($B_1, B_T, B_D$): the linear approximation is valid insofar as only  terms linear in the $B$'s can be safely  retained.  We shall  distinguish between photon absorptions that decrease the atom's kinetic energy and photon absorptions that increase it; we shall also find that, on the average, the spontaneous emission of a photon increases the atom's kinetic energy by an amount equal to the recoil energy $E_R$ (in the linear approximation).  The balance between these competing processes (cooling and heating) characterizes the steady state condition in laser cooling.
 \par
 In section \ref{laser}, the formulas of section \ref{linear} are  applied to laser cooling, in the limit of low laser intensity, i.e. in the limit in which the stimulated emission is negligible. It is assumed that the laser photons are red detuned. The average variation of the kinetic energy $<\Delta E_K>$ of an atom, due to the absorption of a laser photon and the subsequent emission of a fluorescence one, is calculated by taking into account the different probability that the absorbed photon belongs to one of the two counter--propagating laser beams.  The condition    $<\Delta E_K>=0$   yields a unique value of the atom's velocity parameter $B_1$ that depends on the detuning parameter $\delta^*$.  This value of $B_1$ can be minimized as a function of the detuning parameter $\delta^*$, thus obtaining the lowest value $B_{1min}$. The absorbed laser photons decrease the kinetic energy of the atoms with velocity parameter larger than $B_1$ (or $B_{1min}$) and increase the kinetic energy of the atoms with velocity parameter smaller than $B_1$ ($B_{1min}$).
 By identifying   $cB_1$ with $v_{rms}$ (or $cB_{1min}$ with $v_{rms}$),
  we  calculate the average kinetic energy $<E_K>=(1/2)Mc^2{B^2_1}$  in a steady state condition, or the lowest attainable average kinetic energy $<E_K>_{min}=(1/2)Mc^2{B^2_{1min}}$.
  \par
  Finally, in the Appendix, the absorption of a counter--propagation photon by an atom is dealt within Newtonian mechanics in order to highlights the basic conceptual differences with the relativistic treatment, in spite of the fact that the two approximated calculations yields the same equations.
\section{Basic formulas\label{basic}}
The first treatment of the absorption/emission of photons by atoms based on energy and linear momentum conservations is due to Schr\"odinger' \cite{erwin}. Schr\"odinger's paper has been ignored by his contemporaries and even nowadays it is not so popular: see, for instance, \cite{dragan, gg1, gg2}. Reasonably, this oblivion  has been due to the deep rooting of the wave description of light in the background physical knowledge, in spite of the fact that, the absorption/emission of light by atoms is a discrete process.
In \cite{erwin}, Schr\"odinger's dealt only with the emission of photons and,  in particular, he never introduced explicitly the energy difference $\Delta E$ between the two levels of the atomic transition. Therefore, in the following, the original treatment by Schr\"odinger is extended to the absorption case (as in \cite{gg1, gg2}) and the  form of the equations is adapted to the problem under study. Schr\"odinger's approach, besides the conservations laws and special relativity, assumes that  the absorption/emission process is instantaneous: more physically, that the duration of the absorption/emission process is much smaller than the lifetimes of the atomic energy levels.
\begin{figure}[htb]
 \centerline{
 \includegraphics[width=6cm]{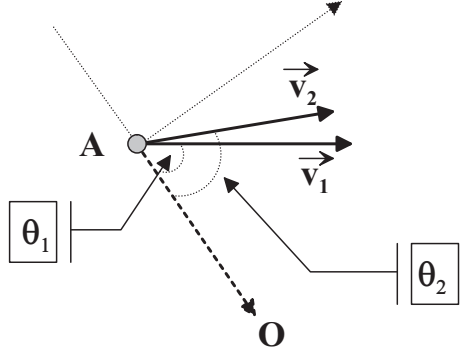}
 }
  \caption{\label{erwinfig}
  Emission of a photon  by the atom $A$ in  motion. The photon is emitted along the direction $A\rightarrow
O$. The subscript $1$ denotes the quantities  before the emission;  the subscript $2$ the quantities after the emission.
  }
 \end{figure}
 \par\noindent
For the emission of a photon by an atom, the conservation laws are (see fig. \ref{erwinfig}):
\begin{equation}\label{energia2}
E_{ph}=\gamma_1 E_1-\gamma_2 E_2
\end{equation}
for energy, and
\begin{eqnarray}
\gamma_1 {{E_1}\over{c^2}}v_1 \cos\theta_1 &= & \gamma_2
{{E_2}\over{c^2}}v_2 \cos\theta_2 +{{E_{ph}}\over{c}}
\label{qmx}\\
\gamma_1 {{E_1}\over{c^2}}v_1 \sin\theta_1 &= & \gamma_2
{{E_2}\over{c^2}}v_2 \sin\theta_2\label{qmy}
\end{eqnarray}
for linear momentum.
$E_{ph}$ is the energy of the emitted photon; $E_1$ and $E_2$  are the rest energies of the atom before and after the emission;  $\gamma_1,\,\gamma_2$ are the relativistic factors before and after the emission; $v_1$ and $v_2$ the atom's velocities before and after the emission; $\theta_1$ and $\theta_2$ the angles between $\vec v_1$ and $\vec v_2$ and the direction of the emitted photon. Notice that $E_1-E_2=\Delta E$, where $\Delta E$ is the energy difference between the two  levels of the atomic transition: $\Delta E$ is a relativistic invariant, since it is given by the difference of two rest energies.
After some calculations, we get:
\begin{equation}\label{interab}
    E_{ph}=\frac{1}{2}\left( \frac{E_1}{\varphi_1}-\frac{E_2}{\varphi_2} \right)
\end{equation}
where:
\begin{equation}\label{varphi}
    \varphi_i= \gamma_i\left(1- \frac{v_i}{c} \cos\theta_i \right)=\gamma_i\left(1- B_i   \cos\theta_i\right); \qquad i=1,2
\end{equation}
Equation (\ref{interab}), by taking into account that $E_1\varphi_1=E_2\varphi_2$, can be written as:
\begin{equation}\label{inter12}
    E_{ph}=\frac{1}{2}\frac{E_1^2-E_2^2}{E_1\varphi_1}=\frac{1}{2}\frac{E_1^2-E_2^2}{E_2\varphi_2}
\end{equation}
Notice that $E_2=Mc^2$ and $E_1= Mc^2+\Delta E$ with $\Delta E\ll Mc^2$. From now on, we shall use the dimensionless parameter $B_T=\Delta E/(2Mc^2)$: as we shall see, $B_T$ is a threshold parameter and the velocity $v_T=cB_T$ a threshold velocity.
\par
It is easy to verify that :
\begin{equation}
 \frac{E_1^2-E_2^2}{2E_2}=  \frac{E_1^2-E_2^2}{2Mc^2}= \Delta E \left(1+\frac{\Delta E}{2Mc^2}  \right)=\Delta E(1+B_T)
 \end{equation}
 \begin{equation}
\frac{E_1^2-E_2^2}{2E_1}  = \Delta E \left(1-\frac{\Delta E}{2E_1}  \right)\approx \Delta E(1-B_T)\label{bta}
\end{equation}
$\Delta E/(2E_1)$ differs from $B_T$ by a term of the order of $B_T^2$; therefore, we shall replace in (\ref{bta}) the $\approx$ sign with that of the equality.
Therefore, from equation (\ref{inter12}), we get:
 \begin{equation}\label{uguale}
E_{ph} = \Delta E \left( 1 - B_T\right) {{\sqrt{1-B_1^2}}\over{1 -B_1\cos \theta_1 }}=\Delta E \left( 1 + B_T\right) {{\sqrt{1-B_2^2}}\over{1 -B_2\cos \theta_2 }}
\end{equation}
or, in  compact form:
\begin{equation}\label{semplice}
    E_{ph}=\frac{\Delta E}{\varphi_1}(1-B_T)=\frac{\Delta E}{\varphi_2}(1+B_T)
\end{equation}
 The case of absorption can be treated in the same way, starting from adequately re-written conservation equations. It turns out that the energy of the absorbed  photon  is given by:
 \begin{equation}\label{ugualeass}
E_{ph} = \Delta E \left( 1 + B_T\right) {{\sqrt{1-B_1^2}}\over{1 -B_1\cos \theta_1 }}=\Delta E \left( 1 - B_T\right) {{\sqrt{1-B_2^2}}\over{1 -B_2\cos \theta_2 }}
\end{equation}
 or by the compact equation:
\begin{equation}\label{sempliceass}
    E_{ph}=\frac{\Delta E}{\varphi_1}(1+B_T)=\frac{\Delta E}{\varphi_2}(1-B_T)
\end{equation}
Both equations (\ref{uguale}) yield the energy of the emitted photon, one in terms of  the atom's velocity parameters before the emission ($B_1, \theta_1$), the other in terms of the atom's velocity parameters after the emission ($B_2, \theta_2$). This last equation, on the other hand, yields the energy of a photon absorbed by an atom with initial velocity parameters ($B_2, \theta_2$) (first equation of (\ref{ugualeass})). Therefore: if an excited atom with velocity parameters  ($B_1, \theta_1$) emits a photon, the same atom, after the emission and, therefore, with velocity parameters  ($B_2, \theta_2$), can absorb a photon of the same energy.
\par
The variation of the atom's kinetic energy due to the emission of a photon is given by:
\begin{equation}\label{emicine}
    \Delta E_K^{emission}=(\gamma_2E_2-E_2)-(\gamma_1E_1-E_1)= \Delta E-E_{ph}
\end{equation}
When the atom is at rest before emission, $\Delta E_K=\Delta E^2/2Mc^2$: by definition, this is the recoil energy $E_R$.
The equation:
\begin{equation}\label{recvel}
    E_R=\frac{1}{2}Mv_R^2
\end{equation}
defines the recoil velocity $v_R=\Delta E/Mc= 2v_T$.
Similarly, the variation of the atom's kinetic energy due to the absorption of a photon is given by:
\begin{equation}\label{asscine}
     \Delta E_K^{absorption}=(\gamma_2E_2-E_2)-(\gamma_1E_1-E_1)= E_{ph}-\Delta E
\end{equation}
The recoil energy in the case of absorption is the same as that in the case of emission.
\section{The natural linewidth: implications\label{naturalsec}}
Each atomic transition has a natural linewidth $\Gamma$ defined as the Full Width  at Half Maximum of a Lorentzian function centered at the  value of  the transition energy $\Delta E$ corresponding to the highest transition probability: we shall denote this value as $\Delta E_M$. The existence of a natural linewidth  suggests to define another  dimensionless  parameter:  $B_D= \Gamma/\Delta E_M$;  $B_D$ defines the Doppler limit in laser cooling since $B_D\Delta E_M(1\pm B_T)\approx B_D\Delta E_M$ describes the first order Doppler effect.
\par
In  laser cooling, we are particularly interested in the absorption of  photons belonging to two counter--propagating laser beams. An atom with velocity parameter $B_1$ can absorb a counter--propagating photon if its transition energy $\Delta E^A$ satisfies the first of equations (\ref{ugualeass}) (the superscript $A$ stays for `anti--parallel'). The same atom can absorb a co--propagating photon if its transition energy $\Delta E^P$ satisfies the same  equation (the superscript $P$ stays for `parallel').
Then:
\begin{equation}\label{viavai}
    \Delta E^P=  \Delta E^A\frac{1-B_1}{1+B_1}
\end{equation}
$\Delta E^P$ is always smaller than $\Delta E^A$ for $B_1\neq 0$ and equal to $\Delta E^A$ for $B_1=0$.
\section{The linear approximation\label{linear}}
In the linear approximation, the energy of the emitted photon is obtained from  equations  (\ref{uguale}) with the approximation $\sqrt{1-B_1^2}\approx 1-B_1^2/2$ and by keeping only  terms linear in  the $B$'s ($B_T$, $B_D$, $B_1$):
\begin{equation}\label{emiapp}
    E_{ph}= \Delta E(1-B_T+B_1\cos\theta_1)=\Delta E(1+B_T+B_2\cos\theta_2)
\end{equation}
 Similarly, the energy  of the absorbed photon, in the linear approximation, is obtained from  equations (\ref{ugualeass}):
\begin{equation}\label{assapp}
    E_{ph}= \Delta E(1+B_T+B_1\cos \theta_1)= \Delta E(1-B_T+B_2\cos \theta_2)
\end{equation}
In the linear approximation, the energy of the emitted/absorbed photon depends only on the atom's velocity component along the direction of propagation of the emitted/absorbed photon.
\subsection{Variation of the atom's kinetic energy: absorption\label{linassappsec}}
As we have seen in equation (\ref{asscine}), the variation of the kinetic energy of the atom due to the absorption of a photon is given by:
\begin{equation}\label{var_kin}
    \Delta E_{K}=  E_{ph}-\Delta E
\end{equation}
Let us suppose that a photon with energy $E_{ph}$ is propagating along the negative direction of the $x$ axis. This photon can be absorbed by any atom whose velocity component $v_x=-v_1\cos\theta_1$ satisfies the first of equations (\ref{assapp}): the variation of the atom's kinetic energy is the same for all these atoms. This means that, in the calculations, we can consider only the cases $\theta_1=\pi$ and $\theta_1=0$.
 Then:
 \begin{equation}\label{cine+-}
    \Delta E_{K}= \Delta E(B_T\pm B_1)
 \end{equation}
where the minus sign corresponds to $\theta_1=\pi$ and the plus sign to $\theta_1=0$.
 \par
 We define a red detuned photon as a photon with energy
$E_{ph}=\Delta E_M(1+B_T)(1-\delta^*)$ where $\Delta E_M(1+B_T)$ is the maximum energy of an absorbed photon when the atom is at rest before absorption and $\delta^*$ is of the same order of magnitude of or smaller than $B_D$: then, $E_{ph}\approx\Delta E_M(1+B_T-\delta^*)$.
 In the linear approximation, this photon can be absorbed in a head on collision  if (first equation of (\ref{assapp})):
 \begin{equation}\label{headonmp}
    \Delta E^A= \Delta E_M(1-\delta^*+B_1 )
 \end{equation}
 The  variation of the atom's kinetic energy  is given by:
 \begin{equation}\label{cinefine}
    \Delta E_K^A= \Delta E_M(B_T-B_1)
  \end{equation}
  If $B_1>B_T$, the absorption decreases the atom kinetic energy: in laser cooling, this is the cooling mechanism. If $B_1<B_T$, the absorption increases the atom kinetic energy; in laser cooling, this is one of the heating processes at work.
        $B_T$ is a threshold parameter.
 The  threshold velocity is:
\begin{equation}\label{soglia}
   v_T=\frac{\Delta E_M}{2Mc}=\frac{v_R}{2}
\end{equation}
         The same atom with the same velocity parameter can absorb a co--propagating red detuned photon of the sa can be absorbed by atoms flying in the same direction of the photon. In this case, the transition energy of the  atom must satisfy the equation:
    \begin{equation}\label{same}
    \Delta E^P= \Delta E_M(1-\delta^*-B_1 )
 \end{equation}
 and the variation of the atom's kinetic energy due to the absorption is:
 \begin{equation}\label{samecin}
    \Delta E_K^P= \Delta E_M(B_T+B_1)
  \end{equation}
When the atom is flying in the same direction of the photon, the variation of the kinetic energy due to the absorption of a photon is always positive. In laser cooling, this is a second heating process.
 \subsection{Variation of the atom's kinetic energy: emission\label{emission}}
 In the case of emission the energy conservation implies that:
 \begin{equation}\label{ene_emi}
    \Delta E_{K}= \Delta E- E_{ph}
 \end{equation}
  According to equation (\ref{emiapp}), the energy of the emitted photon, in the linear approximation, is given by:
 \begin{equation}\label{emiMP}
    E_{ph}= \Delta E(1-B_T+ B_1 \cos\theta_1)
 \end{equation}
 Correspondingly, the variation of the atom's kinetic energy is:
 \begin{equation}\label{pinco}
    \Delta E_K= \Delta E(B_T-B_1\cos\theta_1)
 \end{equation}
 We see that, also for emission, $B_T$ operates as a threshold parameter.
\par
Given an atom with a velocity parameter $B_1$,  it is useful to consider the average energy of the emitted photon under the hypothesis that  any direction of emission is equally probable. From (\ref{emiapp}) we get:
\begin{eqnarray}
   <E_{ph}>&=& \Delta E (1-B_T) +\Delta E B_1\frac{1}{4\pi}\int_0^\pi \cos\theta_1 (2\pi\sin \theta_1d\theta_1)\nonumber\\
   &=& \Delta E (1-B_T)+\frac{B_1\Delta E}{4}\left[\sin^2\theta_1\right]_0^\pi=\Delta E (1-B_T)\label{fotomedia}
\end{eqnarray}
i.e., in the  linear approximation, the average energy of the emitted photon is equal to that of the photon emitted when the emitting atom is at rest before emission.
\par
The average variation of the atom's kinetic energy due to the emission of a photon with the same probability along any direction is given by:
\begin{equation}\label{emimedia}
    <\Delta E_K>=\Delta E- <E_{ph}>=B_T\Delta E\approx B_T\Delta E_M=E_R
\end{equation}
i.e., the average variation of the atom's kinetic energy due to the emission of a photon along an arbitrary direction is positive, independent from $B_1$, and equal to the recoil energy (in the linear approximation). This is a third heating mechanism in laser cooling.
\par
We shall now consider a cycle composed by the absorption of a photon  followed by a spontaneous emission along an arbitrary direction, taking into account the complications due to the linewidth. If an atom absorbs a photon in a head on collision, its transition energy is $\Delta E^A$ and the variation of its kinetic energy is $\Delta E_M(B_T-B_1)$ (equation (\ref{cinefine})).  On the other hand, the average variation of its kinetic energy due to the emission of a fluorescence photon is simply $E_R$. Then, the average overall variation of  its kinetic energy due to the cycle considered is given by:
 \begin{equation}\label{somma}
    <\Delta E_K^A>= \Delta E_M(B_T-B_1)+E_R=\Delta E_M(2B_T-B_1)
 \end{equation}
 This variation  is negative for $B_1>2B_T$, null for $B_1=2B_T$ and positive for $B_1<2B_T$, where $B_1$ is the velocity parameter of the atom before absorption.
 \par
 Similarly, if an atom  flying in the same direction of the photon absorbs it and subsequently undergoes a spontaneous emission, the average variation of its kinetic energy is given by (equations \ref{samecin}, \ref{emimedia}):
 \begin{equation}\label{somma2}
    <\Delta E_K^{P}>= \Delta E_M(2B_T+B_1)
 \end{equation}
 These two last equations contain all the information necessary for a quantitative treatment of laser cooling: taken together, along with the different transition probabilities for $\Delta E^A$ and $\Delta E^P$, they describe all the cooling and heating processes, under the conditions specified in next section.
\section{Laser cooling of two levels atoms\label{laser}}
Before proceeding, it is worth recallig what are the cooling and  heating mechanisms at work. If we assume that the laser photons are propagating along the negative direction  of the $x$ axis, then:
\begin{itemize}
  \item if the atom's velocity component $v_x>v_R/2$, the absorption of a photon decreases the atom's kinetic energy (cooling mechanism);
  \item  if the atom's velocity component $v_x<v_R/2$, the absorption of a photon increases the atom's kinetic energy (heating mechanism);
  \item the emission of a fluorescence photon increases, on the average, the atom's kinetic energy by an amount equal to the recoil energy $E_R$ (heating mechanism).
\end{itemize}
  We shall assume that: the laser photons are mono--energetic;  the laser intensity is low enough so that  stimulated emission is negligible; only the first fluorescence cycle is relevant (this means that the probability that a photon emitted by an atom  is absorbed by another atom is negligible).
\par
   We suppose that the sample of atoms is illuminated by two opposite laser beams of red detuned photons for each direction axis. For symmetry reasons, we can deal only with the two beams propagating, say, along the $x$ direction.
   As stated before, the energy of the red detuned photon is written as:
  \begin{equation}\label{Eph}
    E_{ph}=\Delta E_M(1+B_T)(1-\delta^*)\approx \Delta E_M(1+B_T -\delta^*)
  \end{equation}
If an atom with velocity parameter $B_1$ absorbs a laser photon, this photon belongs to one of the two opposite  beams. If the photon belongs to the  beam flying against the atom, the transition energy $\Delta E^A$  satisfies
  equation (\ref{headonmp}) (in the linear approximation):
\begin{equation}\label{headonmp2}
    \Delta E^A=\Delta E_M(1-\delta^*)  +B_1 \Delta E_M
    \end{equation}
The smallest possible value of $\Delta E^A$ is $\Delta E^A=\Delta E_M(1-\delta^*)$, corresponding to $B_1=0$.
Instead, if the photon belongs to the beam flying in the same direction of the atom, the transition energy  $\Delta E^P$  satisfies
 the equation:
 \begin{equation}\label{viavai2}
    \Delta E^{P}=\Delta E_M(1-\delta^*)-B_1 \Delta E_M=\Delta E^A(1-2B_1)
\end{equation}
The maximum possible value of $\Delta E^P$ is $\Delta E^P=\Delta E_M(1-\delta^*)$, corresponding to $B_1=0$ and equal to the minimum value of $\Delta E^A$.
\par
The transition probabilities $P^A$ and $P^P$  for $\Delta E^A$ and $\Delta E^P$ are different, and given by the corresponding values of the normalized  Lorentzian function describing the natural line.
   The  average variation   of the kinetic energy of the atom due to the absorption of a photon and the subsequent emission of a fluorescence one, weighed  by the relative probabilityted,  is calculated by using  the basic    equations (\ref{somma}, \ref{somma2}) that we reproduce here for convenience:
   \begin{equation}\label{s1}
    <\Delta E_K^{A}> = \Delta E_M(2B_T-B_1)P^{A}
   \end{equation}
   for a photon flying against the atom, and
   \begin{equation}\label{s2}
    <\Delta E_K^{P}> =  \Delta E_M(2B_T+B_1)P^P
   \end{equation}
   for a photon flying in the same direction of the atom.
Therefore, the  average variation of the kinetic energy of the atom  is obtained by summing the two equations (\ref{s1}, \ref{s2}) member by member:
\begin{equation}\label{sommatotale}
     <\Delta E_K>=\Delta E_M[2B_T(P^{A}+P^{P})-B_1(P^{A}-P^{P})]
\end{equation}
Putting:
\begin{equation}\label{put}
    Q(\delta^*)=2B_T\frac{P^A+P^P}{P^A-P^P}
\end{equation}
$<\Delta E_K>$ will be negative if $B_1>Q(\delta^*)$, null if $B_1=Q(\delta^*)$ and positive if $B_1<Q(\delta^*)$.
The condition $B_1=Q(\delta^*)$ yields:
\begin{equation}\label{bello}
 B_1=2B_T\frac{P^A+P^P}{P^A-P^P}
\end{equation}
i.e.
\begin{equation}\label{first}
    B_1=2B_T \frac{L[\Delta E_M(1-\delta^*+B_1)]+L[\Delta E_M(1-\delta^*-B_1)]}{L[\Delta E_M(1-\delta^*+B_1)]-L[\Delta E_M(1-\delta^*-B_1)]}
\end{equation}
where the $L$ is the normalized Lorentzian function describing the natural line shape. After some manipulations we get:
\begin{equation}\label{second}
    B_1^2=\frac{B_T(B_D^2+4\delta^{*^2})}{4(\delta^*-B_T)}
\end{equation}
with $\delta^*>B_T$.  This condition means that the energy of the red detuned photon $E_{ph}=\Delta E_M(1+B_T-\delta^*)$ must be smaller than $\Delta E_M$.  Then:
\begin{equation}\label{detuning}
   {B_1}=\frac{1}{2}\sqrt{B_T \frac{B_D^2+4\delta^{*^2}}{\delta^* -B_T}}
\end{equation}
 The laser beams will reduce the average kinetic energy of the atoms with velocity parameter larger than $B_1$ and will increase the average kinetic energy of the atoms with  velocity parameter smaller than $B_1$.
The smallest value  $B_{1min}$    of $B_1$ is obtained for
\begin{equation}\label{delta*}
    \delta^*=B_T+\frac{1}{2}\sqrt{4B_T^2+B_D^2}
\end{equation}
 and
is given by:
\begin{equation}\label{bminimo}
{B_{1min}}={B_T}\sqrt{2+2\sqrt{1+\frac{B_D^2}{4B_T^2}}}
\end{equation}
  These  results are valid for any value of the ratio $B_D/B_T=\Gamma/E_R$.
 It is interesting to consider three limits: $B_D\gg B_T$, $B_D=B_T$ and $B_D\ll B_T$. In the first case,  $B_{1mim}$ is obtained for $\delta^*\approx B_D$; in the second, for $\delta^*=2.41\, B_T$; in the third for $\delta^* \approx 2B_T$.
\par
\begin{figure}[h]
  \centering{
  \includegraphics[width=10cm]{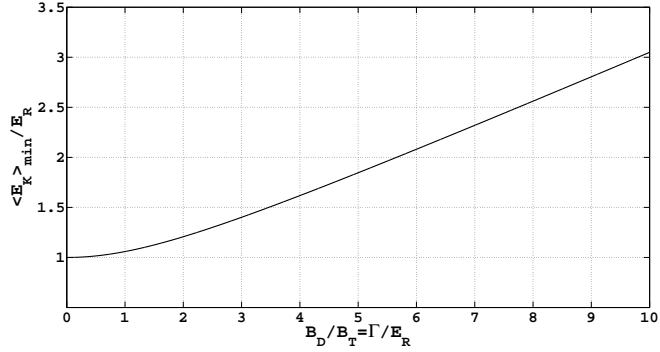}\\
  }
  \caption{The lowest average kinetic energy goes to $E_R$ as the linewidth $\Gamma$ goes to zero.}\label{minimo}
\end{figure}

The absorption of a laser photon depends only on the atom's velocity component $v_x$.
 Since the statistical distribution of $v_x$ is symmetric around $v_x=0$, the atoms' sample in the steady state can be described as if all the atoms have velocity component $v_x=\pm v_{x_{rms}}$. When this velocity is equal to $\pm cB_1$, the average kinetic energy of the atoms can not be reduced further. Then, by  putting $cB_1=v_{x_{rms}}$, we get:
  \begin{equation}\label{rms}
  <E_K>=\frac{1}{2} M<v_x^2>=\frac{1}{2}Mc^2B_1^2=\frac{1}{16}E_R\frac{B_D^2+{4\delta^*}^2}{B_T(\delta^*-B_T)}
 \end{equation}
  The lowest attainable kinetic energy will be:
 \begin{equation}\label{lowest}
    <E_K>_{min}=\frac{1}{2}Mc^2B_{1min}^2=\frac{1}{2}E_R \left(1+\sqrt{1+\frac{B_D^2}{4B_T^2}}  \right)
 \end{equation}
 This equation implies that $<E_K>_{min}\rightarrow E_R$ as $B_D/B_T\rightarrow 0$, i.e as the natural linewidth $\Gamma\rightarrow 0$ (Fig. \ref{minimo}).
\par\noindent
When $\Gamma =0$, red detuned photons can be absorbed only by atoms flying against the photons and the minimum condition  $<E_K>_{min}= E_R$ can be derived directly from equation (\ref{somma}). In fact, according to (\ref{somma}), the average variation of the atom's kinetic energy due to the absorption of a photon and the subsequent emission of a fluorescence one is zero for $B_1=2B_T$. Then, the average kinetic energy of an atom in the steady state condition in given by:
\begin{equation}\label{zerogamma}
    <E_K>=\frac{1}{2}Mc^2B_1^2=\frac{1}{2}Mc^2(2B_T)^2=E_R
\end{equation}
   \subsection{Comparison with the semiclassical theory}
 From equations (\ref{rms}, \ref{lowest}), we get, dividing member by member:
 \begin{equation}\label{ratio}
    \frac{<E_K>}{<E_K>_{min}}=\frac{1}{8}\frac{B_D^2+{4\delta^*}^2}{(\delta^*-B_T)(B_T+\frac{1}{2}\sqrt{4B_T^2+B_D^2})}
 \end{equation}
If $B_T=0$, this equation  reduces to:
\begin{equation}\label{semi}
    \frac{<E_K>}{<E_K>_{min}} =\frac{1}{2}\left(\frac{B_D}{2\delta^*}+\frac{2\delta^*}{B_D}\right)
\end{equation}
In literature, the detuning parameter $\delta$ is defined as $\delta=\omega_l-\omega_a$ where $\omega_l$ and $\omega_a$ are the laser and the atomic transition frequencies. Therefore: $\delta^*=\hbar|\delta|/\Delta E_M$.
Taking into account this relation and that $\Gamma=\hbar\gamma$ (where $\gamma$ is the natural width expressed in terms of angular frequency),  equation (\ref{semi}) assumes the form:
\begin{equation}\label{classica}
 \frac{<E_K>}{<E_K>_{min}}=\frac{1}{2}\left(\frac{\gamma}{2|\delta|} + \frac{2|\delta|}{\gamma}\right)
\end{equation}
 This is the result of the semiclassical theory  of two levels atoms (at low laser intensities): it is a limiting case of the present treatment based on conservation laws.
\par
 As a matter of fact, if we approximate equations (\ref{rms}) and (\ref{lowest}) for $\delta^*\approx B_D\gg B_T$, we obtain:
\begin{equation}\label{cinappcl}
    <E_k>=\frac{\Gamma}{8}\left( \frac{B_D}{2\delta^*} + \frac{2\delta^*}{B_D}\right)=\frac{\hbar\gamma}{8}\left( \frac{\gamma}{2|\delta|}  +\frac{2|\delta|}{\gamma}  \right)
\end{equation}
and
\begin{equation}\label{cinappclm}
    <E_K>_{min}=\frac{\hbar\gamma}{4}
\end{equation}
respectively. We recognize in equation (\ref{cinappcl}) the average kinetic energy of the atom  and in equation (\ref{cinappclm}) the lowest attainable average kinetic energy predicted by the semiclassical theory.
\par
\begin{figure}[h]
  \centering{
      \includegraphics[width=10cm]{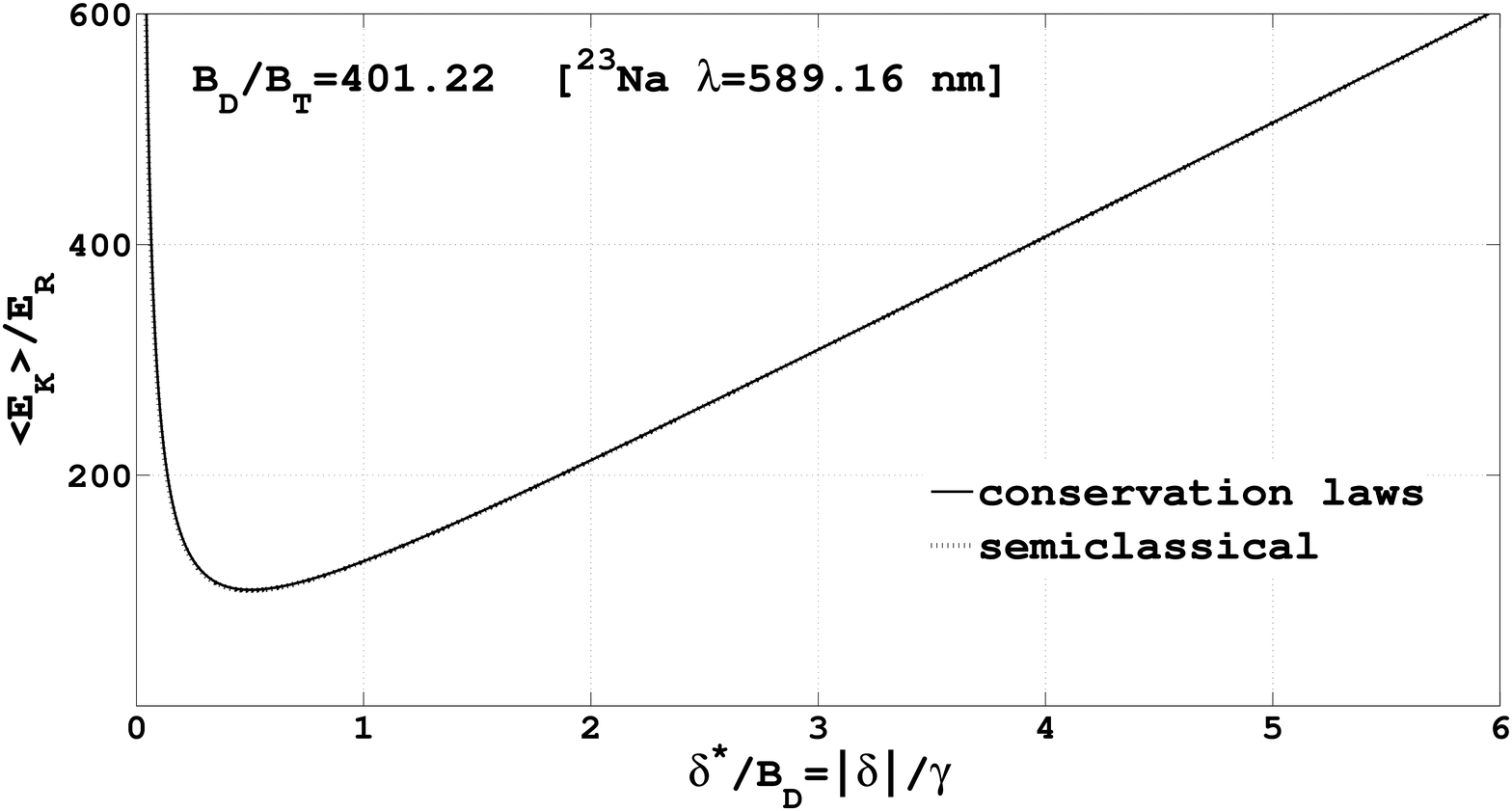}\\
      }
    \caption{Comparison  between the semiclassical theory (dotted line) and the present treatment for the case of the $589.16$ nm transition of $^{23} Na$. }\label{sodio}
\end{figure}

The comparison  with the semiclassical theory can be  visualized by three figures.
  Fig. \ref{sodio} shows that the two treatments are practically indistinguishable when $B_D\gg B_T$.
\par
\begin{figure}[h]
  \centering{
   \includegraphics[width=10cm]{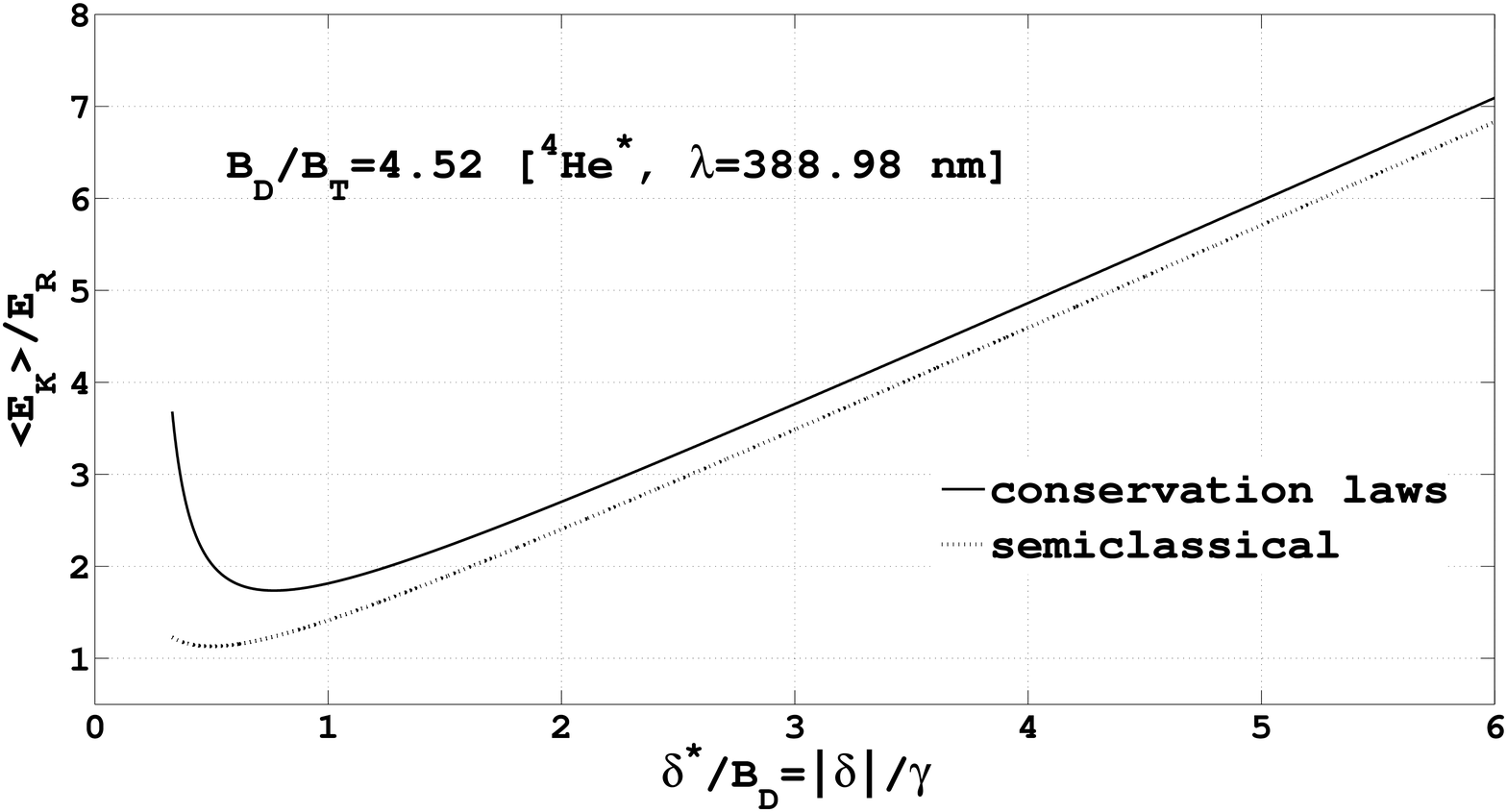}\\
   }
    \caption{Comparison  between the semiclassical theory (dotted line) and the present treatment for the case of the $388.98$ nm transition of $^{4} He^*$. The $^*$ remind us that the starting level of the transition is a metastable one. }\label{elio}
\end{figure}

\noindent
In the semiclassical theory, it is explicitly assumed that the photon momentum  is negligible with respect to the atomic one. This approximation implies that the   kinetic energy of the atom is much larger than the recoil energy. Consequently, in the semiclassical theory both the average kinetic energy and the lowest attainable average kinetic energy  depend only on $\Gamma$ (on $B_D$ in the language of this paper): equations (\ref{cinappcl}) and (\ref{cinappclm}).
Instead, in the present treatment, they depend on both $B_D$ and $B_T$  (equations (\ref{rms}, \ref{lowest})).
However, when $B_D\approx B_T$ the differences between the two treatments are evident (fig. \ref{elio}).
\par
\begin{figure}[h]
\centering{
    \includegraphics[width=10cm]{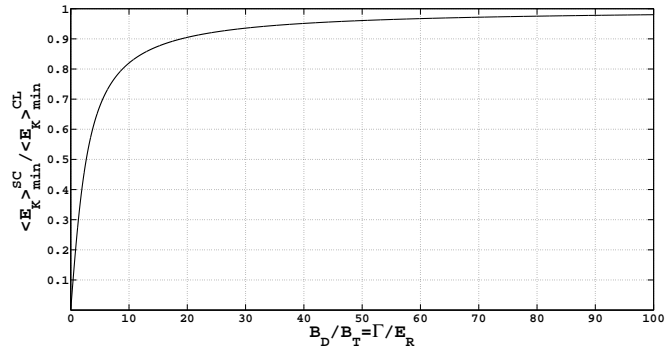}\\
    }
    \caption{Ratio between the  lowest values of the average kinetic energy predicted by the semiclassical theory $<E_K>_{min}^{SC}$ and the conservation laws $<E_K>_{min}^{CL}$.} \label{tmtd}
\end{figure}

\noindent
Finally, the lowest attainable kinetic energy is systematically lower  in the semiclassical theory (Fig. \ref{tmtd}):
the ratio between the value predicted by  the semiclassical theory and the one obtained by conservation laws  tends to one for large values of the ratio $B_D/B_T$ but drops
 dramatically as $B_D/B_T\rightarrow 0$. Of course, this behavior is due to the fact that $B_T$ does not enter into the semiclassical theory.
 \section{Comparison with quantum mechanical treatments\label{qm}}
 Quantum mechanical treatments of laser cooling of two levels atoms at low laser intensities have been carried out by several authors. Wineland and Itano \cite{wineland} deal with both free and bound atoms and begin with formulas based on conservation laws applied to the absorption/emission of photons by atoms.   The  difference with the present treatment lies in the fact that Wineland and Itano's calculation of the steady state average kinetic energy requires the knowledge of the atoms' velocity distribution, assumed to be always Gaussian. For free atoms, when the linewidth is much smaller than the recoil energy, they found that it should be possible to achieve average kinetic energies lower than the recoil energy. However, Wineland and Itano stress that it is difficult to obtain ``the proper conditions under
which these results hold  \cite[p. 1525]{wineland}.''
Instead, in the present paper, the calculation of the average kinetic energy is based on the position $cB_1=v_{x_{rms}}$ where $cB_1$ is the atom's velocity which zeroes the average kinetic energy variation due to an absorption--emission cycle: no knowledge of the velocity distribution is required. As a consequence, as shown above,  when $\Gamma\rightarrow 0$ $<E_K>\rightarrow E_R$: $E_R$ is the lowest kinetic energy allowed by conservation principles in laser cooling of two levels atoms at low laser intensities.
In the second part of their paper, Wineland and Itano show that the quantum mechanical treatment of free atoms does not alter the picture given in the first part of their paper.
\par
The quantum mechanical treatment by Castin, Wallis and Dalibard does not make any assumption on the atoms' velocity distribution  and is valid for any value of the ratio between the natural linewidth and the recoil energy. As in the present paper, they show that the semiclassical theory is valid as long as the natural linewidth is much larger than the recoil energy; however, differently from the present paper, they found that, for very narrow transition lines, the lowest attainable average kinetic energy is about $0.5 E_R$ instead of $E_R$. It is not clear to me why a quantum mechanical treatment  yields this limit value which seems to be incompatible with the conservation laws.
     \section{Conclusions}
The straightforward application of energy and linear momentum conservation to the absorption/emission of photons by atoms allows to find out the essential features of laser cooling of two levels atoms at low laser intensities.
The lowest attainable average kinetic energy of the atoms  depends on the ratio $\Gamma/E_R$ between the natural linewidth and the recoil energy and tends to $E_R$  as $\Gamma/E_R$ tends to zero. This  treatment, like the quantum mechanical ones, is valid for any value of the ratio between the natural linewidth and the recoil energy and contains the results of the standard semiclassical theory of laser cooling as the limiting case in which the recoil energy is negligible with respect to the natural linewidth.

 \renewcommand{\theequation}{A-\arabic{equation}}
  % redefine the command that creates the equation no.
  \setcounter{equation}{0}  % reset counter
  \section*{Appendix: Absorption of a photon in Newtonian mechanics \label{newton}}  % use *-form to suppress numbering
Suppose that an atom,  flying in the positive direction of the $x$ axis, absorbs a counter--propagating photon with energy $E_{ph}$.
The conservation of linear momentum reads:
\begin{equation}\label{momentum}
    Mv_1-\frac{E_{ph}}{c}=Mv_2
\end{equation}
where $v_1$ and $v_2$ are the component of the atom velocity along the $x$ axis, before and after the absorption, respectively. Notice how the light speed $c$, extraneous to classical mechanics, enters  this equation through the photon linear momentum.
The conservation of energy reads:
\begin{equation}\label{energy}
    \frac{1}{2}Mv_1^2+ E_{ph}=\frac{1}{2}Mv_2^2+\Delta E
\end{equation}
where $\Delta E$ is the energy difference between the two levels of the atomic transition.
From (\ref{momentum}):
\begin{equation}\label{velocity}
    v_2=v_1-\frac{E_{ph}}{Mc}
\end{equation}
Substituting this value in (\ref{energy}), we get:
\begin{equation}\label{energy2}
    E_{ph}= \Delta E+\frac{1}{2Mc^2}E_{ph}^2-B_1E_{ph}
\end{equation}
where $B_1=v_1/c$.
If we write:
\begin{equation}\label{position}
    E_{ph}=\Delta E(1+\alpha)
\end{equation}
equation (\ref{energy2}) becomes:
\begin{equation}\label{energy3}
    \Delta E(1+\alpha)=\Delta E +\frac{\Delta E}{2Mc^2}\Delta E(1+\alpha)^2 -B_1\Delta E(1+\alpha)
\end{equation}
Putting $B_T=\Delta E/2Mc^2$ and retaining only terms of the first degree in $\alpha, B_T, B_1$ (linear approximation) we obtain:
\begin{equation}\label{finale}
    \alpha= B_T-B_1
\end{equation}
Therefore:
\begin{equation}\label{finale2}
    E_{ph}=\Delta E(1+B_T-B_1)
\end{equation}
which is identical to the first of equations (\ref{assapp}) in the case of $\theta_1=\pi$.
From (\ref{energy}) and (\ref{finale2}), the variation of the atom's kinetic energy is obtained:
\begin{equation}\label{cinetica}
    \Delta E_K= E_{ph}-\Delta E=\Delta E(B_T-B_1)
\end{equation}
If $B_1=0$, $E_{ph}=\Delta E(1+B_T)$ and $\Delta E_K=\Delta E^2/2Mc^2=E_R$. $E_R$ is the recoil energy.
\par
The rest energy of special relativity $Mc^2$ appears in these equations as a consequence of having considered the absorption of a photon. Of course, in classical mechanics, the physical meaning of $Mc^2$ remains obscure.
\par
The energy of an emitted can be derived in a similar way, by writing down the adequately re--written conservation equations.

\vskip5mm\par\noindent
{\bf Acknowledgments.}
 Thanks are due to Biagio Buonaura, for having sympathetically and carefully followed the development of this paper from the beginning; to Nicole Fabbri, for her critical reading of a first draft of the paper and for her encouragement; and to Ennio Arimondo, whose stringent  comments have  been a kind of driving force for the attainment  of the final version of the paper.

\end{document}